\documentclass[%
 reprint,
 amsmath,amssymb,
 aps,
]{revtex4-1}

\usepackage{graphicx}
\usepackage{dcolumn}
\usepackage{bm}

\begin{document}


\title{\boldmath Hints for a Low $B_s\to \mu^+\mu^-$ Rate and the Fourth Generation}

\author{Wei-Shu Hou, Masaya Kohda, and Fanrong Xu}
 \affiliation{
  Department of Physics, National Taiwan University,Taipei, Taiwan 10617 
}


\begin{abstract}
With full 2011 LHC data analyzed, there is no indication
for deviation from Standard Model (SM) in CP violating phase 
for $B_s \to J/\psi\phi$, nor in the forward--backward 
asymmetry for $B^0\to K^{*0}\mu^+\mu^-$.
SM sensitivity, however, has been reached for 
$B_s \to \mu^+\mu^-$ rate, and there 
may be some hint for a suppression.
We illustrate that, if a suppressed 
${\cal B}(B_s \to \mu^+\mu^-)$ bears out with 2012 data,
it would imply a lower bound on the 
fourth generation quark mixing product $|V_{t's}^*V_{t'b}|$.
\begin{description}
\item[PACS numbers]
14.65.Jk 
12.15.Hh 
11.30.Er 
13.20.He 
\end{description}
\end{abstract}

\pacs{Valid PACS appear here}
\maketitle


\section{\label{sec:Intro}INTRODUCTION\protect\\}

The Winter conferences have brought forth
a host of new experimental results from the LHC.
Continuing the 2011 trend, the Standard Model (SM)
stands tall, and there are no strong
hints of new physics beyond SM (BSM).
On the flavor front, a fit~\cite{psiphiLHCb}
to $B_s \to J/\psi\phi$ events by LHCb with
1 fb$^{-1}$ data yields $\Delta\Gamma_s$ that is
in good agreement with SM, while
combining the $\phi_s \equiv 2\Phi_{B_s}$ (the $CP$
violating phase of $\bar B_s \to B_s$ mixing)
measurement with the result from $B_s\to J/\psi\pi\pi$
gives
\begin{eqnarray}
\phi_s &=& -0.002 \pm 0.083 \pm 0.027\ {\rm rad}
  \ \ \ {\rm (LHCb\ 1\ fb}^{-1}) \nonumber \\
 &=& -0.002 \pm 0.087\ {\rm rad},
\label{phis1203}
\end{eqnarray}
which is fully consistent with the result of
$0.03 \pm 0.16 \pm 0.07$ with 1/3 the dataset.
Again with 1 fb$^{-1}$ data, LHCb has advanced
the measurement of forward-backward asymmetry
in $B^0 \to K^{*0}\mu^+\mu^-$, giving a first
measurement~\cite{s0LHCb} of the zero-crossing point
\begin{equation}
 q_0^2 = (4.9^{+1.1}_{-1.3})\ {\rm GeV}^2,\ \ \ \
  {\rm (LHCb\ 1\ fb}^{-1})
\label{s01203}
\end{equation}
which is consistent with SM expectation of 
4.0--4.3 GeV$^2$~\cite{Bobeth}.

It is interesting then, that more apparent progress
has been made on the quest for the $B_s \to \mu^+\mu^-$
rare decay mode: SM sensitivity has genuinely been
reached, and data~\cite{mumuCMS, mumuLHCb} might be
suggestive of a rate {\it below} SM expectations.
Given that a decade long search for $B_s \to \mu^+\mu^-$
was motivated by the possible enhancement up to factors
of hundreds to thousands, by powers~\cite{tanbn}
of $\tan\beta$ in the settings of supersymmetry 
or two Higgs doublet models, 
we are now at the
juncture of a mindset change, switching from possible
huge enhancements of old,
to SM-like or even sub-SM values as it might emerge.
It is in this context that we wish to explore in
this Brief Report the implications on relevant flavor
parameters involving a fourth generation of quarks,
$t'$ and $b'$ (SM4).

It should be noted that bounds on $t'$ and $b'$ masses
have reached~\cite{4G-MoriQCD} the 600 GeV level by direct search at the
LHC, hence we have nominally crossed the threshold of
the unitarity bound (UB) of 500--550 GeV~\cite{Chanowitz:1978uj}.
In the following, we will proceed naively, extending
our previous work~\cite{HKX}, and return to comment on
UB and other issues towards the end of our discussion.

\section{\label{sec:II}\boldmath 
Low versus SM-like $B_s \to \mu^+\mu^-$ Rate \protect\\}

It is difficult to enhance $B_s \to \mu^+\mu^-$ in SM4
by more than a factor of two or three,
because it is constrained by $B \to X_s\ell^+\ell^-$
(together with $B \to X_s\gamma$), which is consistent with SM.
Hence, this mode appeared less relevant for SM4, until recently.
In contrast, the aforementioned $\tan\beta$ enhancement effect
feeds scalar operators that do not enter
$b\to s\gamma$ and $b\to s\ell^+\ell^-$ processes,
hence were far less constrained.
However, the scalar operators are now muted by
the prowess of the LHC (and previous searches at the Tevatron).

A dramatic turn of events were already played out in 2011,
where the combined result~\cite{comboBsmumu} of LHCb and CMS,
 ${\cal B}(B_s\to \mu^+\mu^-) < 11 \times 10^{-9}$ at
 95\% Confidence Level (CL),
refuted the CDF result~\cite{CDFmumu11} of
$(18^{+11}_{-\ 9}) \times 10^{-9}$, which was
at the time itself hot-off-the-press.
Adding close to 3 fb$^{-1}$ data to the previous 7 fb$^{-1}$
analysis, the CDF value dropped a bit to
$(13^{+9}_{-7}) \times 10^{-9}$, but
the Tevatron has ran out of steam.
ATLAS has also turned out a bound of $22 \times 10^{-9}$
based on 2.4 fb$^{-1}$ data, which is not yet competitive
even with summer 2011 results from LHCb or CMS.
The highlight this Winter was therefore the
$B_s\to \mu^+\mu^-$ results from CMS and LHCb.

Let us first describe the LHCb result.
Using a multivariate analysis (MVA),
LHCb gave~\cite{mumuLHCb} the 95\% CL bound of
\begin{equation}
 {\cal B}(B_s\to \mu^+\mu^-) < 4.5 \times 10^{-9},\ \ \
  {\rm (LHCb\ 1\ fb}^{-1})
\label{LHCb-mumu-1fb}
\end{equation}
which is approaching rather close to the SM value~\cite{Buras:2010wr} of
\begin{equation}
 {\cal B}(B_s\to \mu^+\mu^-) = (3.2 \pm 0.2) \times 10^{-9}.\ \ \
  {\rm (SM)}
\label{mumu-SM}
\end{equation}
In fact, LHCb gave a fitted number,
\begin{equation}
 {\cal B}(B_s\to \mu^+\mu^-) = (0.8^{+1.8}_{-1.3}) \times 10^{-9},\ \
  {\rm (LHCb\ 1\ fb}^{-1})
\label{mumu-subSM}
\end{equation}
which naively implies possibly negative branching ratio!
The central value is from the maximum log-likelihood,
while the errors correspond to varying the
log-likelihood by 0.5. The main upshot may be that
LHCb does not really see any clear hint of a
SM-strength signal!
Either this is a downward fluctuation of the ``true SM"
value of Eq.~(\ref{mumu-SM}), or Nature has
a sub-SM value in store for us.
We caution, of course, that statistics is still rather low.

The CMS result~\cite{mumuCMS} is, at 95\% CL,
\begin{equation}
 {\cal B}(B_s\to \mu^+\mu^-) < 7.7 \times 10^{-9},\ \ \
  {\rm (CMS\ 5\ fb}^{-1})
\label{CMS-mumu-5fb}
\end{equation}
by a cut-based analysis. A mild deficit seems
to be indicated when compared with the
median expected limit of $< 8.4 \times 10^{-9}$.
But the handful of events reveal some
interesting pattern.
In the Barrel detector region, one expects
$\sim 2.7$ signal events if SM were true,
together with $\sim 0.8$ events from background.
Only two events were observed, which are
separated by $\sim 100$ MeV, wider than the
nominal detector mass resolution.
This suggests the presence of background events.
Whether this constitutes one event each for signal
and background, or if both events are background,
it seems to echo LHCb~\cite{mumuLHCb} in some ``downward"
fluctuation from the SM value of Eq.~(\ref{mumu-SM}).
However, if both LHCb and CMS sense a downward signal
fluctuation, then the likelihood that the actual
signal might be lower would be enhanced!

In the Endcap detector region, the situation is
a bit puzzling. Here, signal and background are
both expected at 1.2 event level, while a total of
4 events were seen~\cite{mumuCMS}.
But they all cluster within 50 MeV or less,
inside a signal mass window of 150 MeV,
which is set at twice the detector mass resolution
(poorer than in the Barrel detector).
However, since the Endcap is less sensitive
than the Barrel, we refrain from further comment,
except that the ``excess" events push up the
bound of Eq.~(\ref{CMS-mumu-5fb}) slightly.
Thus, by CMS Barrel detector alone, the
``discrepancy" with median expected is
a little larger.

Although anything can happen at the present
statistics level, LHCb expects to add
$\sim$ 1 fb$^{-1}$ in 2012, while
CMS would add $\sim$ 15 fb$^{-1}$, both at
the slightly higher collision energy of 8 TeV.
We therefore like to emulate
future prospects as follows.
For the indication of lower than SM rate,
we shall take Eq.~(\ref{mumu-subSM}) at face
value. Projecting to full 2011-2012 data,
besides the doubling of LHCb data, CMS data should
increase more than four fold (an MVA approach
should increase the effective luminosity).
Although one cannot really project what is
the combined effective reduction of errors,
we take the factor $\sqrt{6} \sim 2.5$.
I.e. in our subsequent numerics, besides the
$1\sigma$ allowed range for Eq.~(\ref{mumu-subSM}),
we will show also the $1/2.5\, \sigma$ range,
which would give $(0.8^{+0.7}_{-0.5}) \times 10^{-9}$.
While this is rather aggressive, it would
illustrate a sub-SM result when LHCb combined
with CMS probes genuinely below SM values.
It is not impossible that, by end of
2011-2012 run, we find ${\cal B}(B_s\to \mu^+\mu^-)$
to be consistent with zero, i.e. at $10^{-9}$ or less.
We note that ATLAS could also eventually contribute
significantly to $B_s\to \mu^+\mu^-$ search.

The notable feature across the board for
new physics search at the LHC, however, is
that no cracks were found in SM's armor.
Thus, we offer a second case of SM-like behavior.
Here, we take the central value from SM,
and mimic the current error bar by satisfying
the 95\% CL bound from CMS. We get from
Eq.~(\ref{CMS-mumu-5fb}),
\begin{equation}
 {\cal B}(B_s\to \mu^+\mu^-) = (3.2 \pm 2.7) \times 10^{-9}.\ \ \
  \textrm{(SM-like)}
\label{mumu-SMlike}
\end{equation}
Again, we will discuss the $1\sigma$ and $1/2.5\, \sigma$
allowed range of Eq.~(\ref{mumu-SMlike}) for projections
into the future.
Actual error reduction would likely be more than
$1/2.5$ for SM-like central values in Eq.~(\ref{mumu-SMlike}).

We follow our previous paper~\cite{HKX} and combine the above
scenarios for ${\cal B}(B_s\to \mu^+\mu^-)$ with
measurements of $\phi_s$ and
$A_{\rm FB}(B^0 \to K^{*0}\mu^+\mu^-)$
(we shorthand as $A_{\rm FB}$ below).
Our target physics is the flavor parameters of the
fourth generation for $b\to s$ transitions, namely
$V_{t's}^*V_{t'b} \equiv r_{sb}\, e^{i\,\phi_{sb}}$.
If the current hint for 125 GeV SM-like Higgs boson
does not get substantiated by 2012 data, a very
heavy fourth generation could provide the mechanism
for electroweak symmetry breaking through its strong
Yukawa interaction~\cite{Hou:2012nh}.
We will find that a sub-SM ${\cal B}(B_s\to \mu^+\mu^-)$
value would imply a \textit{lower bound} on
$r_{sb} = \vert V_{t's}^*V_{t'b}\vert$,
which would be rather interesting.

We had suggested that the
three measurements of $\phi_s$,
${\cal B}(B_s\to \mu^+\mu^-)$ and $A_{\rm FB}$
would help map out the preferred $V_{t's}^*V_{t'b}$,
or $(r_{sb},\ \phi_{sb})$ parameter space.
The main measurements are $\phi_s$ and
${\cal B}(B_s\to \mu^+\mu^-)$, with $A_{\rm FB}$
providing further discrimination, both in its
shape, and now also the $q_0^2$ value~\cite{s0LHCb}.
Three cases were discussed.
Case A was for large and negative $\phi_s$,
where we used $\sin2\Phi_{B_s} = -0.3 \pm 0.1$,
and enhanced $10^9{\cal B}(B_s\to \mu^+\mu^-)
= 5.0 \pm 1.5$. This was motivated by
hints for large and negative time-dependent
CPV in $B_s \to J/\psi \phi$ from Tevatron studies.
Although a $-0.2 \pm 0.1$ value could still be
entertained at the 2$\sigma$ level, there is
not more to be said beyond our previous work,
while the likelihood for enhanced
${\cal B}(B_s\to \mu^+\mu^-)$ is receding.
Thus, we no longer present this case.
Case B and C were for $\sin2\Phi_{B_s}$
taking SM value of $-0.04 \pm 0.01$,
while $10^9{\cal B}(B_s\to \mu^+\mu^-)$
takes the slightly enhanced or depressed values
of $5.0 \pm 1.5$ and $2.0 \pm 1.5$, respectively.
By design, the overlap between Case B and Case C
is precisely when ${\cal B}(B_s\to \mu^+\mu^-)$
is SM-like.
Thus, the three Cases of A, B and C map out
the foreseen parameter space in $r_{sb}$
and $\phi_{sb}$ as data improves.

With the present experimental situation
for ${\cal B}(B_s\to \mu^+\mu^-)$,
which could either be sub-SM as in
Eq.~(\ref{mumu-subSM}), or SM-like, as in
Eq.~(\ref{mumu-SMlike}), we reinvestigate
the implications for the preferred region
in the $r_{sb}$-$\phi_{sb}$ plane.
For both cases, we impose the
$\phi_s \equiv 2\Phi_{B_s}$ constraint of
Eq.~(\ref{phis1203}).
The observed shape and $q_0^2$ value from
$A_{\rm FB}$ are further applied to
constrain parameter space.
We take $m_{t'} = 650$ GeV for sake of illustration.

\begin{figure*}[t!]
\centering
{\includegraphics[width=70mm]{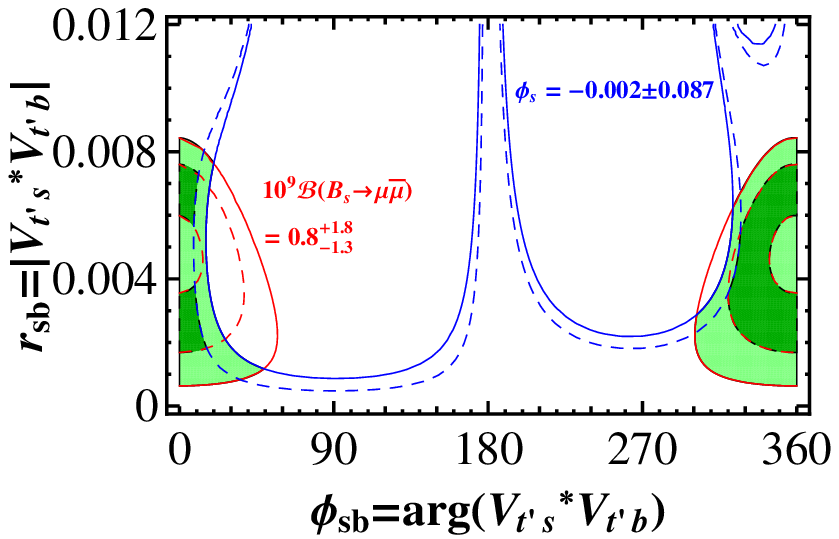}
 \includegraphics[width=70mm]{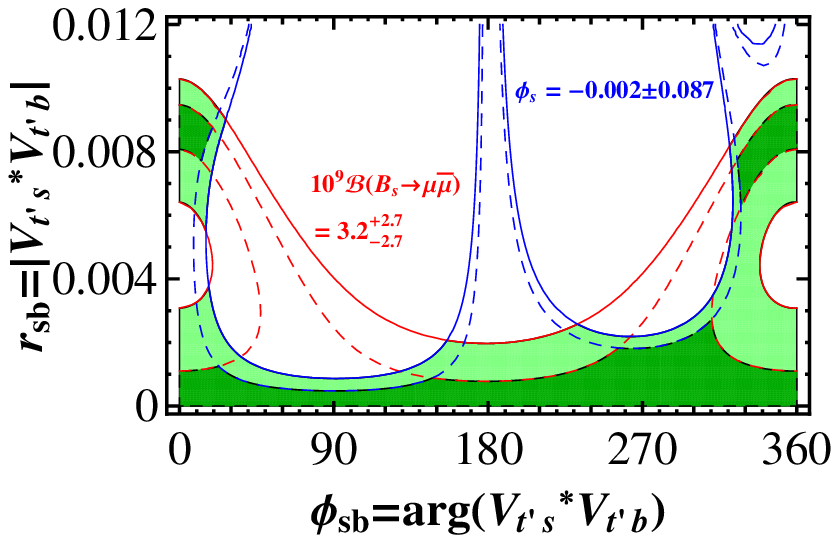}
}
\vskip-0.35cm
\caption{
 Overlap region for contours of
  $\phi_{s} = -0.002 \pm 0.087$,
   where dashed line is for $1/\sqrt{2}$ the error, and
  $10^{9} {\cal B}(B_s \to \mu^+\mu^-) = $
  (a) $0.8^{+1.8}_{-1.3}$
  (we allow only positive definite values), and
  (b) $3.2 \pm 2.7$,
  where dashed line is for $1/2.5$ the error.
  For illustration, $m_{t'} = 650$ GeV has been used.
} \label{Fig1}
\end{figure*}

\section{\boldmath 
Results \protect\\}

The $\bar B_s$--$B_s$ mixing amplitude is
\begin{eqnarray}
 M_{12}^s &=& \frac{G_F^2M_W^2}{12\pi^2}m_{B_s}f_{B_s}^2\hat B_{B_s}\eta_B
   \Delta_{12}^s,
 \label{M12s}
\end{eqnarray}
with
\begin{equation}
\Delta_{12}^s = \left(\lambda_t^{\rm\scriptsize SM}\right)^2 S_0(t,t)
    + 2\lambda_t^{\rm\scriptsize SM}\lambda_{t'}\Delta S_0^{(1)}
    + \lambda_{t'}^2\Delta S_0^{(2)},
 \label{Del12s}
\end{equation}
where $\lambda_q \equiv V_{qs}^*V_{qb}$.
With $S_0$ and $\Delta S_0^{(i)}$ as defined in Ref.~\cite{Hou:2006mx},
Eq.~(\ref{Del12s}) manifestly respects GIM~\cite{Glashow:1970gm}.
The CPV phase
\begin{equation}
 \phi_s = 2\Phi_{B_s} \equiv \arg M_{12}^s = \arg \Delta_{12}^s,
 \label{argM12s}
\end{equation}
depends only on $m_{t'}$ and $\lambda_{t'} = V_{t's}^*V_{t'b}$.
Note that $\lambda_t^{\rm\scriptsize SM} =
-\lambda_c - \lambda_u \cong -0.04 -V_{us}^*V_{ub}$.
Although we take PDG~\cite{PDG} values for the
phase of $V_{ub}$, it is exciting that the
phase of $V_{us}^*V_{ub}$ is starting to be
directly measured via interference of tree processes.
For ${\cal B}(B_s \to \mu^+\mu^-)$, the $f_{B_s}^2$ dependence is
largely removed~\cite{Buras:2003td} by taking the
ratio with $\Delta m_{B_s}/\Delta m_{B_s}|^{\rm exp}$,
which works for SM4 as in SM. That is,
\begin{equation}
\mathcal{B}(B_s\to \bar\mu\mu)
 = C\frac{\tau_{B_s}\eta_Y^2}{\hat{B}_{B_s}\eta_B}
   \frac{|\lambda_t^{\rm\scriptsize SM}Y_0(x_t)+\lambda_{t'}\Delta Y_0|^2}
        {|\Delta_{12}^s|/\Delta  m_{B_s}|^{\rm exp}},
 \label{BrBsmumu}
\end{equation}
where $C = 3g_W^4m_{\mu}^2/2^7\pi^3M_{W}^2$, and
$\eta_Y= \eta_Y(x_t) = \eta_Y(x_{t'})$ is taken.

We plot, in Fig.~1(a), the contours for $\phi_{s}$
within $1\sigma$ and $1/\sqrt{2}\,\sigma$ range of
Eq.~(\ref{phis1203}), in the
$r_{sb} \equiv |V_{t's}^*V_{t'b}|$,
$\phi_{sb} \equiv \arg V_{t's}^*V_{t'b}$
plane, for $m_{t'} = 650$ GeV.
Here, LHCb holds a monopoly, and statistics is
expected to only double during 2012.
Similarly for ${\cal B}(B_s\to \mu^+\mu^-)$,
we plot the contours within $1\sigma$ and $1/2.5\,\sigma$
range of Eq.~(\ref{mumu-subSM}), which is sub-SM
in strength.
The $m_{t'}$ value used is beyond the 550 GeV
nominal UB bound~\cite{Chanowitz:1978uj},
and one is no longer sure of the
numerical accuracy of Eqs.~(\ref{Del12s}) and (\ref{BrBsmumu}),
i.e. the perturbative computation of the functions
$\Delta S_0^{(i)}$ and $\Delta Y_0$ would become questionable.
However, some form such as Eq.~(\ref{Del12s}) should
continue to hold even above the UB,
and we shall continue to use existing formulas.

The overlap between the $\phi_{s}$ and
${\cal B}(B_s\to \mu^+\mu^-)$ contours
now favor $\phi_{sb}$ in the 4th quadrant
with $|\sin\phi_{sb}|$ small,
where the darker regions are for more
aggressive error projections towards the future.
It should be clear that a precise determination of
$\phi_{sb}$ depends much more on the precision of
$\phi_{s}$ measurement.

We remark that Fig.~1(a) is a much more
stringent version of Case C presented
in our previous paper, where we now have
${\cal B}(B_s\to \mu^+\mu^-)$ considerably
below SM expectation.
Thus, the most notable feature is that,
even at $1\sigma$ error level,
$r_{sb}$ is now \textit{bounded from below}.
This is because the (current LHCb~\cite{mumuLHCb})
central value of Eq.~(\ref{mumu-subSM}) is
more than $1\sigma$ below the SM expectation of
$3.2\times 10^{-9}$. Thus, it calls for a finite
$t'$ effect to subtract, or destructively interfere,
against the SM amplitude from top quark. That this
might become the picture for flavor parameters involving
4th generation, if a lower than SM value for
${\cal B}(B_s\to \mu^+\mu^-)$ is found at the LHC,
is the main point of this short note.
It should be stressed that this is a natural
consequence for SM4, since we know from
existing constraints that $t'$-induced amplitudes
must be subdominant in strength compared with
top-induced amplitudes, while the
sign of the real part of $V_{t's}^*V_{t'b}$
can precisely be correlated with what experiments
observe.

The SM-like case of Eq.~(\ref{mumu-SMlike}) is
less interesting, but given the continued
success of the SM into the LHC era,
should be viewed as more probable.
We illustrate in Fig.~1(b) for $m_{t'} = 650$ GeV
the overlap of the contours for $\phi_{s}$
in Eq.~(\ref{phis1203}) and
${\cal B}(B_s \to \mu^+\mu^-)$
in Eq.~(\ref{mumu-SMlike}).
Besides some high $r_{sb}$ region for
modest $\vert \phi_s\vert$ values, the generic
feature is relatively small $r_{sb}$, with
$\phi_{sb}$ undetermined by the present
precision of $\phi_{sb}$ measurement.
This small $r_{sb}$ case is rather intuitive,
that of subdued 4th generation effect.
We shall see that the larger $r_{sb}$
values are ruled out by the observation of
SM-like behavior for $A_{\rm FB}$,
as we have seen in our previous paper.

The SM-like shape for $A_{\rm FB}$ as observed by
LHCb provides a powerful discriminant
against larger $r_{sb}$ values.
Note that data prior to summer 2011 had
suggested a deviation from SM behavior~\cite{PDG},
which, besides a hint for sizable
deviation in $\sin2\Phi_{B_s}$, was part of
the motivation for Case A in our previous paper.
The SM-like shape for $A_{\rm FB}$ is further
affirmed with 1 fb$^{-1}$ data from LHCb~\cite{s0LHCb},
while the first measurement for
zero-crossing point, Eq.~(\ref{s01203}), is offered.
We have checked the allowed parameter space
of Fig.~1 and find generically
that $r_{sb} \gtrsim 0.004$ would generate
significant deviations in shape for $A_{\rm FB}$.
The drop from roughly $0.008$~\cite{HKX} to $0.004$
is due to the higher $m_{t'} = 650$ GeV taken
to satisfy direct search bounds~\cite{4G-MoriQCD},
as well as the tighter experimental constraints
towards SM.
We note with interest that, for the sub-SM
${\cal B}(B_s \to \mu^+\mu^-)$ case,
the slightly larger than SM central value of
$q_0^2 = 4.9$ GeV$^2$ in Eq.~(\ref{s01203})
also prefers $\phi_{sb}$ in the 4th quadrant.

\section{\label{sec:Conclusion} Discussion and Conclusion\protect\\}

After some hints for BSM physics for some years,
both in $A_{\rm FB}(B^0 \to K^{*0}\mu^+\mu^-)$
and in $\sin2\Phi_{B_s}$~\cite{disc},
SM is reaffirmed by 2011 data from LHC.
Interestingly, now there might be a hint for
${\cal B}(B_s \to \mu^+\mu^-)$ below SM expectations.
It is of course too early to tell. However,
this mode has always been looked upon as
possibly greatly enhanced by
the less constrained scalar operators.
We are at least at the turning point,
where no large enhancement is observed,
but now whether it is SM-like, or sub-SM,
can be distinguished with full 2011-2012 data.
The 4th generation $t'$ quark offers the
natural toolbox in this domain,
as it is constrained to be subdominant
by $b\to s\gamma$ and $b \to s\ell^+\ell^-$
data since a decade,
while providing a destructive mechanism
in the unknown phase of $V_{t's}^*V_{t'b}$.
In contrast,
adjusting the scalar interactions to the SM strength
is like training a big hammer on a small nail.

We note that, to have ${\cal B}(B_s \to \mu^+\mu^-)$
near the central value of Eq.~(\ref{mumu-subSM}),
the $C_{10}$ Wilson coefficient would be
considerably smaller than SM value,
such that one would worry about ${\cal B}(B \to X_s\ell^+\ell^-)$.
It is then interesting to not that
LHCb data does seem to indicate that the
$d{\cal B}(B^0 \to K^{*0}\mu^+\mu^-)/dq^2$ differential rate
could be a little lower than the SM expectations~\cite{s0LHCb}.
Although a vanishing ${\cal B}(B_s \to \mu^+\mu^-)$ is unlikely
(more probably within $(1-2)\times 10^{-9}$),
it would be interesting to watch this mutually supporting
trend of somewhat lower ${\cal B}(B_s \to \mu^+\mu^-)$
and ${\cal B}(B \to K^*\ell^+\ell^-)$ 
(or ${\cal B}(B \to X_s\ell^+\ell^-)$).
The darker region in Fig.~1(a) is just to stress the point.

We have used $m_{t'} = 650$ GeV to satisfy direct
search bounds, which is now beyond the nominal
unitarity bound. To probe much further,
the 13-14 TeV run would be necessary.
However, with the Yukawa coupling turning nonperturbative,
the phenomenology may change~\cite{Enkhbat:2011vp}.
Fortunately, the leading production mode of
$gg\to Q\bar Q$ is not affected.
The usage of such large $m_{t'}$ values is becoming
dubious, and nonperturbative studies should be
performed.
The nonperturbative, strong Yukawa coupling
could actually be the source of electroweak symmetry
breaking~\cite{Hou:2012nh}.
It is interesting that, with full 2011-2012 data,
we would learn whether a 125 GeV Higgs boson is
substantiated, as well as whether $B_s\to \mu^+\mu^-$
is below SM expectations.

In conclusion, we illustrate what LHC data
might tell us about 4th generation flavor parameters.
Assuming 2011-2012 data would give
$\phi_s = -0.002 \pm 0.062$ and taking $m_{t'} = 650$ GeV,
we mocked the low $B_s \to \mu^+\mu^-$ rate case with
$(0.8^{+0.7}_{-0.5}) \times 10^{-9}$, which would imply
$|V_{t's}^*V_{t'b}| \sim 0.0015$--0.004, with
$-40^\circ \lesssim \arg V_{t's}^*V_{t'b} \lesssim 15^\circ$.
On the other hand, if a SM-like $B_s \to \mu^+\mu^-$ rate
emerges, it would imply small $|V_{t's}^*V_{t'b}|$
at a couple per mille, while $\arg V_{t's}^*V_{t'b}$
would require more precise measurement of $\phi_s$
to determine.
The $B^0 \to K^{*0}\mu^+\mu^-$ forward-backward
asymmetry provides a further discriminant that rules out
$|V_{t's}^*V_{t'b}| \gtrsim 0.004$ for the discussed
allowed regions.

\vskip0.3cm
\noindent{\bf Acknowledgement}.
WSH is grateful to
U. Langenegger, S. Stone and D.~Tonelli
for communications,
and thanks the National Science Council for
the Academic Summit grant NSC 100-2745-M-002-002-ASP.
MK and FX are supported by the NTU grant
10R40044 and the Laurel program.

\end{document}